# Phase lagging model of brain response to external stimuli - modeling of single action potential

Karthik Seetharaman[1], Hamidreza Namazi[2], Vladimir V. Kulish[3]

[1] Research Unit of Neuropysiology and Biomechanics of Movement, Universite Libre de Bruxelles, Brussels, Belgium

[2] School of Mechanical and Aerospace Engineering, Nanyang Technological University, Singapore
Email: m080012@e.ntu.edu.sg   Tel: +6583740288   (Corresponding Author)

[3] School of Mechanical and Aerospace Engineering, Nanyang Technological University, Singapore

**ABSTRACT**

In this paper we detail a phase lagging model of brain response to external stimuli. The model is derived using the basic laws of physics like conservation of energy law. This model eliminates the paradox of instantaneous propagation of the action potential in the brain. The solution of this model is then presented. The model is further applied in the case of a single neuron and is verified by simulating a single action potential. The results of this modeling are useful not only for the fundamental understanding of single action potential generation, but also they can be applied in case of neuronal interactions where the results can be verified against the real EEG signal.

## 1. Introduction

Brain as the most complex organ in the human body controls all bodies' actions/reactions by receiving different stimuli through the nervous system. Any stimulus stronger than the threshold stimulus is translated by the number of sensory neurons generating information about the stimulus and the frequency of the action potentials. After the action potential has been generated, it travels through the neural network to the brain. In various sections of the network and the brain, integration of the signals takes place. Different areas of the brain respond based on the kind and the location of stimuli. The brain sends out signals which generate the response mechanism.

During many years, numerous models of brain/cortical activity have been proposed by scientists. On a microscopic level, the work that was done by Freeman in the modeling of EEG arising from the olfactory bulb of animals during the perception of odors is noteworthy to be mentioned. He developed a set of non-linear equations for this response which generates EEG like pattern [1-2]. When the microscopic models are extended to a macro level, then different methods are employed. Many of these models assume the cortical region to be a continuum. The first continuum model was proposed by Lopes da Silva et al. and Van Rotterdam et al. In this model, which was proposed by Lopes da Silva et al., two interacting populations of cells were considered namely, the pyramidal cells and the local interneurons. These two populations is characterized by two transfer systems, one representing the excitatory post synaptic potentials (EPSPs), and the other one representing inhibitory post synaptic potentials (IPSPs). The potential of the pyramidal cell dendrites is the signal to which the recorded EEG was assumed to be proportional [3]. Van Rotterdam et al. extended this model by assuming an infinite one-dimensional chain of interconnected pyramidal cells and interneurons. They obtained a transfer function for the temporal frequency and spatial frequency which took into account the effect of inhibitory and excitatory populations of neurons. With suitable adjustments, the model recreated the alpha rhythm. The major issue with this model was that it did not include the nonlinear effects, axonal delays (delays due to finite propagation velocity of signals in the axons of neurons) and convolutions of the cortex [4]. After this Nunez introduced the effect of axonal delays and proposed a model which allowed wave solutions, and identified the alpha rhythm as being at the fundamental cortical Eigenfrequency. Nunez also solved this model for a 1D loop cortex in two-dimensional cortex with some boundary conditions, but this model neglected the inhomogeneity of the cortical connections and the effect of the convolutions of the cortex as in Lopes da Silva's model [5]. Wright and Liley introduced a spatially discretized model which takes into account the nonlinear effects with axonal delays and dendritic delays, but it did not involve the convolutions and non-uniformity of the cortical connections. This model could only be used for small systems or a coarser look at the larger systems [6-7]. Robinson et al. took Wright and Liley's equations and introduced a continuum wave model to replace the linear discrete parts of the equations based on the same fundamentals; they also simplified the computation of dendritic lags, but this model did not consider the filtering effects of the skull on the EEG which they claimed could be solved by using MEG with EEG [8]. Liley et al. developed a set of non-linear continuum field equations which described the macroscopic dynamics of neural activity in the cortical region [9]. These equations were used by Steyn-Ross et al. who introduced noise terms into them to give a set of stochastic partial differential equations (SPDEs). They also converted equations governed by Liley et al. into linearized ODEs. This model could predict the substantial increase in low frequency power at the critical points of induction and emergence. They later used this model to study the electrical activity of an anaesthetized cortex [10-13]. Kramer et al. started with the equations given by Steyn-Ross and co-workers and neglected the spatial variation and the stochastic input. This gave rise to a set of ordinary differential equations (ODEs) for the modeling of cortical activity. Then, they showed that the results obtained from the SPDE model agree with clinical data in an approximate way, but they also stated that the spatial sampling of the cortex was poor because of inherent shortcomings in the equipment used. There is also ambiguity in the lead/lag relationships determined by the Windowed Cross Correlation that they used. They suggested that the seizing activity of the cortex may be understood as an example of pathological pattern formation [14]. Kulish and Chan have suggested a novel method for the modeling of brain response using the fundamental laws of nature like energy conservation and the least action principle. The model equation obtained has been solved and the results show a good agreement with real EEGs [15].

Even though, many researchers have worked on the modeling of brain for years, there is no mathematical model which effectively considers the effect of lagging time into account. The primary reason offered for the absence of such a model is the complexity and non-linearity of the brain.

This paper attempts to suggest a phase-lagged model of single action potential which is based on fundamental laws of physics. In order to consider the least complexity of the human brain here we propose a model in the microscopic level of brain organization which is a single neuron. This model considers a finite time lag between the occurrence of the stimulus and the rise of the subsequent action potential in a node Ranvier which eliminates the paradox of instantaneous propagation of energy.

## 2. Impulse Generation and Propagation Mechanism

In order to study the human behavior and in a casual manner neural activity, one can consider the different level of brain organization at many scales in time and space from a single neuron to the whole brain organization [16].

Microscopic level of brain organization refers to the activity of single-neurons with their dendritic networks [17]. Neurons are the principal elements of the nervous system.

The research data suggests that appropriate environmental stimuli bigger than the threshold value are the main reason of the dendritic fields of individual cortical neurons, to which the neurons are exposed, thus, in a causal manner brain organizations make the proper decision [18-19].

An Axon, or nerve fiber, is a long, thin process of a neuron which typically conducts electrical impulses away from the neuron's cell body. In vertebrates, Axons are enclosed in an insulating myelin sheath formed by special neuroglia cells (Figure. 1). The myelin sheath increases the speed of impulse transmission and prevents the gates on that part of the axon from opening and exchanging their ions with the outside environment.

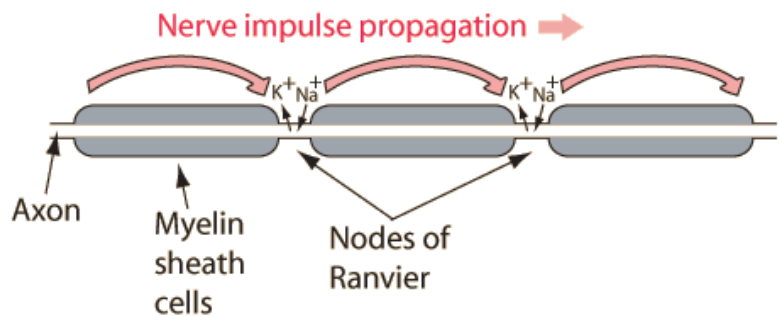

Figure. 1. Schematic structure of a typical axon

There are gaps periodically between the myelin sheath segments known as the Nodes of Ranvier. At those uncovered areas of the axon membrane, the ion exchange necessary for the production of an action potential can take place.

The inactive or resting neuron actively pumps sodium ions ($Na^+$) out the cell and potassium ions ($Ka^+$) into the cell. However, the plasma membrane is more permeable to $Ka^+$ ions than $Na^+$, so $Ka^+$ diffuses out of the neuron which results in an excess of positively charged ions outside the neurons membrane and an excess of negatively charged ions inside the neurons. This unusual distribution of electrical charges on each side of the plasma membrane makes the membrane polarized, and this condition is known as the resting potential with the voltage of -70 mV. Potential is the difference in electrical charge between two sites. The polarization of the neuron membrane does not change as long as the neuron is inactive.

When the neurons stimulated as a result of receiving a stimulus, they exhibit an all-or-none response. They either form an impulse, or they do not respond. The weakest stimulus which activates a neuron is called a threshold stimulus. When a neuron is activated by a stimulus, its plasma membrane instantly becomes permeable to $Na^+$, so these ions quickly diffuse into the neuron. The inward flow of $Na^+$ causes positive and negative ions to be equally abundant on each of the plasma membrane. Thus, there is not net electrical charge on either side of the membrane. The plasma membrane is now depolarized. This sudden depolarization is the nerve impulse, or action potential (Figure. 2). Then, the wave of depolarization flows along the myelinated axon.

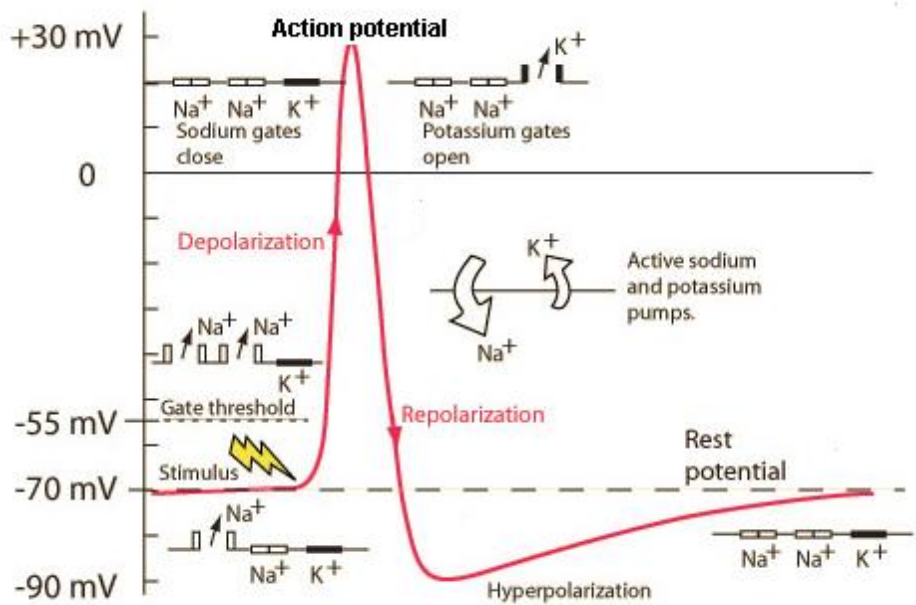

Figure. 2. Action potential generation in a neuron

Immediately after depolarization, $Ka^+$ diffuses outward to reestablish the resting potential of the membrane, with an excess of positive charges outside and an excess of negative charges inside. In this way, the neuron membrane is repolarized. Then, $Na^+$ is pumped out and $Ka^+$ is pumped into the neurons in order to reestablish the resting-state distribution of ions. When this is accomplished, the neuron is ready to respond to another stimulus. Depolarization and Repolarization are accomplished in about 1/1000 of a second each one.

### 3. Diffusion model of voltage change in a neuron

The derivation of this model shall follow the derivations as outlined by Plonsey and Barr [20]. The detailed derivation is presented in their book. A short version of the derivation, for the purpose of understanding, shall be given in this section.

Considering the neuron as a cylindrical conductor, it is assumed that the conductor is axially symmetrical, and the current conducted by that neuron is also axial in nature. This means that the potentials and currents in the neuron are functions of a single variable.

By Ohm's Law, the decrease in potential in the neuron per unit length must be equal to the IR drops along the neuron and are given by:

$$\frac{\partial \varphi_{ext}}{\partial x} = -I_{ext} r_{ext} \tag{1}$$

$$\frac{\partial \varphi_{int}}{\partial x} = -I_{int} r_{int} \tag{2}$$

where $\varphi_{ext}$ is the extracellular potential, $\varphi_{int}$ is the intracellular potential, $I_{ext}$ is the extracellular current, $I_{int}$ is the intracellular current, $r_{ext}$ is the extracellular resistance, and $r_{int}$ is the intracellular resistance.

By Kirchhoff's current law, we have:

$$\frac{\partial I_{int}}{\partial x} = -i_m \tag{3}$$

where $i_m$ is the transmembrane current per unit length. This is in line with the conservation of current law which requires the axial rate of decrease in the intercellular current be equal to transmembrane current per unit length. This is because any loss in the intercellular current has to be due to the current crossing the membrane.

For extracellular current, any decrease in its value is due to current that is crossing the membrane along with the current that is carried out due to the electrodes insertion. So:

$$\frac{\partial I_{ext}}{\partial x} = i_m + i_p \qquad (4)$$

where $i_p$ is the current that is carried outside because of the polarizing electrodes inserted in the membrane. Let $I$ be defined as:

$$I = I_{int} + I_{ext} \qquad (5)$$

The membrane potential $V_m$ is given by:

$$V_m = \varphi_{int} - \varphi_{ext} \qquad (6)$$

differentiating the above equation (6) with respect to $x$ and including equations (1) and (2), we have:

$$\frac{\partial V_m}{\partial x} = \frac{\partial \varphi_{int}}{\partial x} - \frac{\partial \varphi_{ext}}{\partial x}$$

$$= -r_{int} I_{int} + r_{ext} I_{ext}$$

$$= -r_{int} I_{int} + r_{ext}(I - I_{int}) \qquad (7)$$

From equations (3) and (4), we have:

$$\frac{\partial I}{\partial x} = i_p \qquad (8)$$

we also have:

$$\frac{\partial V_m}{\partial x} = -(r_{int} + r_{ext})I_{int} + I r_{ext} \qquad (9)$$

differentiating equation (9) with respect to $x$:

$$\frac{\partial^2 V_m}{\partial x^2} = -(r_{int} + r_{ext})\frac{\partial I_{int}}{\partial x} + r_{ext}\frac{\partial I}{\partial x} \qquad (10)$$

substituting equations (3) and (8):

$$\frac{\partial^2 V_m}{\partial x^2} = (r_{int} + r_{ext})i_m + r_{ext}\, i_p \qquad (11)$$

For an action potential on a single fiber we must have no external polarizing current i.e. $i_p = 0$. Also, from equations (5) and (8), $I_{int} = -I_{ext}$. Under such constraints, the equation (11) becomes:

$$\frac{\partial^2 V_m}{\partial x^2} = (r_{int} + r_{ext})i_m \qquad (12)$$

Also, the nerve is present in an extensive extracellular medium. This makes the extracellular resistance $r_{ext} = 0$. This approximation is also permitted by the Hodgkin-Huxley experimental chamber during their studies on the squid axon. The intracellular resistance per unit length $r_{int}$ can be rewritten in terms of the intracellular resistivity, $R_{int}$ as:

$$r_{int} = \frac{R_{int}}{\pi a_c^2} \tag{13}$$

where $a_c$ is the radius of the cylindrical nerve. So:

$$i_m = \frac{\pi a_c^2}{R_{int}} \frac{\partial^2 V_m}{\partial x^2} \tag{14}$$

The transmembrane current per unit area $I_m$ can also be written as:

$$I_m = \frac{i_m}{2\pi a_c} \tag{15}$$

so:

$$I_m = \frac{a_c}{2 R_{int}} \frac{\partial^2 V_m}{\partial x^2} \tag{16}$$

Equation (16) can be equated to the transmembrane current density from the Hodgkin- Huxley relations to give:

$$\frac{a_c}{2 R_{int}} \frac{\partial^2 V_m}{\partial x^2} = C_m \frac{\partial V_m}{\partial t} + \sum I_{ion} \tag{17}$$

where $C_m$ is the membrane capacitance and $\sum I_{ion}$ is the sum total of the ionic currents. Rearranging equation (17):

$$\frac{\partial^2 V_m}{\partial x^2} = \frac{2 C_m R_{int}}{a_c} \frac{\partial V_m}{\partial t} + \frac{2 R_{int}}{a_c} \sum I_{ion} \tag{18}$$

If the equation (18) is observed, it is noticed that this diffusion equation shall give rise to the paradox of instantaneous propagation of action potential. This means that the equation suggests that the response to the stimulus signal is generated almost instantaneously. This is physically not possible as that would require an infinite speed of propagation. This paradox of instantaneous propagation exists for all such diffusion equations which become evident as the characteristic time gets shorter.

In order to avoid such a paradox, the space time behavior of the action potential should satisfy the wave equation i.e.

$$V_m(x,t) = V_m(x - ct) \tag{19}$$

where, $c$ is the velocity of propagation. Differentiating equation (19) twice and using the chain rule, we get:

$$\frac{\partial^2 V_m}{\partial x^2} = \frac{\partial^2 V_m}{\partial t^2}\frac{1}{c^2} \tag{20}$$

Adding equations (18) and (20), we get:

$$2\frac{\partial^2 V_m}{\partial x^2} = \frac{1}{c^2}\frac{\partial^2 V_m}{\partial t^2} + \frac{2C_m R_{int}}{a_c}\frac{\partial V_m}{\partial t} + \frac{2R_{int}}{a_c}\sum I_{ion} \tag{21}$$

## 4. Phase-lagging model of voltage change in a neuron

The paradox of instantaneous propagation can be rectified by introducing a time lag term which takes care of the lag between the occurrence of the stimulus and the rise of the subsequent action potential. Thus, the constitutive relation to be coupled with the conservation equation must be:

$$\Psi(x, t+\tau) = -D\frac{\partial V}{\partial x}(x,t) \tag{22}$$

where $\frac{\partial V}{\partial x}$ on the right hand side of equation (22) is a direct result of the external influence i.e. stimulus and the term $\tau$ is the time lag which makes sure that the paradox of instantaneous propagation does not come into play.

Thus, the quantities involved in equation (22) are written for two different instances of time. In order to bring those physical quantities at the same time instance, it is necessary to expand the left hand side of equation (22) in Taylor's series, and, keeping in mind the lagging time is small in comparison with the transient time of the process, to neglect those terms of the expansion whose order is larger than one i.e.,

$$\Psi(x,t) + \tau\frac{\partial \Psi}{\partial x}(x,t) \cong -D\frac{\partial V}{\partial x}(x,t) \tag{23}$$

which is the Fick's law with a finite lagging time.

We need to couple the new constitutive relation expressed by equation (23) with the conservation equation. In order to do so, let us find the derivative with respect to $x$ of equation (23). We obtain:

$$\Psi_x + \tau\frac{\partial \Psi_x}{\partial x} \cong -DV_{xx} \tag{24}$$

from the conservation equation it follows that:

$$\Psi_x = -V_x + f(x,t) \tag{25}$$

where $f(x,t)$ is the source function.

Upon substituting (25) into (24), we obtain:

$$\tau\frac{\partial^2 V}{\partial t^2} + \frac{\partial V}{\partial t} = D\frac{\partial^2 V}{\partial x^2} + f(x,t) + \tau f_t(x,t) \tag{26}$$

denoting

$$S(x,t) = f(x,t) + \tau f_t(x,t) \tag{27}$$

We have:

$$\tau \frac{\partial^2 V}{\partial t^2} + \frac{\partial V}{\partial t} = D \frac{\partial^2 V}{\partial x^2} + S(x,t) \tag{28}$$

where $\tau = \frac{D}{c^2}$.

In equation (28) speed of propagation $c$ is a finite quantity which is defined as the speed of impulse traveling through the neuron. Also, The diffusivity term $D$ is related to the resistance of the neuron to the electrical impulse. $D$ is the property of the neural tissue and will dampen the impulse as it travels over the nerve.

The source/sink term $S(x,t)$ arises from the constant interactions between the neurons even in the absence of an external potential. It also accounts for the fact that the voltage is not conserved.

Equation (28) is now solved in a semi-infinite domain. Here, we consider that the axon is long compared to the distance over which signals propagate by passive electrotonic conduction (i.e., a semi-infinite axon). The initial condition can be defined as $V = V_0$ at the time $t = 0$ which is the resting potential of a neuron, and $\frac{\partial T}{\partial t} = 0$.

Three new variables shall be introduced now in equation (28):

$$-\xi = \frac{x}{D^{\frac{1}{2}}}$$

$$\vartheta = \frac{t}{\tau^{\frac{1}{2}}}$$

$$\Lambda = V - V_0 \tag{29}$$

The term $\xi$ is the new distance like variable, $\vartheta$ is the new time like variable and $\xi$ is the excess potential. Now the equation (28) becomes:

$$\frac{\partial^2 \Lambda}{\partial \vartheta^2} + \tau^{-\frac{1}{2}} \frac{\partial \Lambda}{\partial \vartheta} = \frac{\partial^2 \Lambda}{\partial \xi^2} + S(\xi, \vartheta) \tag{30}$$

The initial conditions in this case are $\Lambda = 0$ at $\vartheta = 0$ and $\frac{\partial \Lambda}{\partial \vartheta} = 0$.

Taking the Laplace Transform of equation (30) and rearranging the terms, we get:

$$\frac{d^2 \tilde{\Lambda}}{d\xi^2} - s\left(s + \tau^{-\frac{1}{2}}\right)\tilde{\Lambda} + \tilde{S}(\xi, s) = 0 \tag{31}$$

where $\tilde{\Lambda}(\xi, s)$ is the Laplace transform of the excess voltage and $\tilde{S}(\xi, s)$ is the Laplace transform of the source/sink term.

The general solution of Equation (31) is:

$$\tilde{\Lambda}(\xi, s) = C_1(s)\exp\left\{-\xi\left[s\left(s + \tau^{-\frac{1}{2}}\right)\right]^{\frac{1}{2}}\right\} + C_2(s)\exp\left\{\xi\left[s\left(s + \tau^{-\frac{1}{2}}\right)\right]^{\frac{1}{2}}\right\} + P(\xi, s) \tag{32}$$

The first two terms on the right hand side of the equation (32) are the general solution of the associated homogenous equation, when $\tilde{S}(\xi, s) = 0$, and $P(\xi, s)$ is the particular solution.

Since we are considering a semi-infinite domain, the solution has to be bounded as $\xi \to \infty$ so as to not violate the energy conservation principle. This requires that $C_2(s) \to 0$. Denoting $C_1(s) \equiv C(s)$, we get:

$$\tilde{\Lambda}(\xi, s) = C(s)\exp\left\{-\xi\left[s\left(s + \tau^{-\frac{1}{2}}\right)\right]^{\frac{1}{2}}\right\} + P(\xi, s) \tag{33}$$

Rearranging equation (33), C(s) can be written as:

$$C(s) = [\tilde{\Lambda}(\xi, s) - P(\xi, s)]\exp\left\{\xi\left[s\left(s + \tau^{-\frac{1}{2}}\right)\right]^{\frac{1}{2}}\right\} \tag{34}$$

Equation (34) shall be differentiated with respect to $\xi$ in order to eliminate $C(s)$ as shown by Kulish and Lage [21].

$$\frac{d\tilde{\Lambda}}{d\xi} = -\left[s\left(s + \tau^{-\frac{1}{2}}\right)\right]^{\frac{1}{2}} C(s)\exp\left\{-\xi\left[s\left(s + \tau^{-\frac{1}{2}}\right)\right]^{\frac{1}{2}}\right\} + \frac{dP(\xi,s)}{d\xi}$$

$$= -\left[s\left(s + \tau^{-\frac{1}{2}}\right)\right]^{\frac{1}{2}}\tilde{\Lambda}(\xi, s) + \left[s\left(s + \tau^{-\frac{1}{2}}\right)\right]^{\frac{1}{2}} P(\xi, s) + \frac{dP(\xi,s)}{d\xi}$$

$$-\tilde{\Lambda}(\xi, s) = \left[s\left(s + \tau^{-\frac{1}{2}}\right)\right]^{-\frac{1}{2}}\frac{d\tilde{\Lambda}}{d\xi} - \left[s\left(s + \tau^{-\frac{1}{2}}\right)\right]^{-\frac{1}{2}}\frac{dP(\xi,s)}{d\xi} - P(\xi, s) \tag{35}$$

From Abramowitz and Stegun (page 374-379, 1024) [22], it is found that the inverse Laplace Transform of $\left[s\left(s + \tau^{-\frac{1}{2}}\right)\right]^{-\frac{1}{2}}$ is $\mathfrak{I}_0\left(\frac{\theta}{2\tau^{\frac{1}{2}}}\right)\exp\left(-\frac{\theta}{2\tau^{\frac{1}{2}}}\right)$ where $\mathfrak{I}_0(z)$ is the modified Bessel function.

Thus, taking the Laplace Transform of equation (35) and applying the convolution theorem, we obtain:

$$\Lambda = -\int_0^\theta \frac{\partial \Lambda}{\partial \xi}\mathfrak{I}_0\left(\frac{\theta-\xi}{2\sqrt{\tau}}\right)\exp\left(-\frac{\theta-\xi}{2\sqrt{\tau}}\right)d\xi + \int_0^\theta \frac{dP(\xi,\theta)}{d\xi}\mathfrak{I}_0\left(\frac{\theta-\xi}{2\sqrt{\tau}}\right)\exp\left(-\frac{\theta-\xi}{2\sqrt{\tau}}\right)d\xi + P(\xi, \theta) \tag{36}$$

where $p(\xi,\vartheta)$ is the inverse Laplace transform of the function $P(\xi,s)$.

Restoring the original terms and rearranging the terms, equation (36) transforms into:

$$V(x,t) = V_0 - \left(\frac{D}{\tau}\right)^{\frac{1}{2}} \int_0^t \frac{\partial V}{\partial x} \mathfrak{I}_0\left(\frac{t-t^*}{2\tau}\right) \exp\left(-\frac{t-t^*}{2\tau}\right) dt^* + \left(\frac{D}{\tau}\right)^{\frac{1}{2}} \int_0^t \frac{\partial P}{\partial x} \mathfrak{I}_0\left(\frac{t-t^*}{2\tau}\right) \exp\left(-\frac{t-t^*}{2\tau}\right) dt^* + P(x,t) \quad (37)$$

Equation (37) gives the relation between the voltage and its spatial derivative at any moment in time and at any location in the domain of interest.

It is noteworthy to mention that because the length of a node of Ranvier, where a single action potential generated, is in the range of 1 μm long, thus, here we don't analyze the spatial component "x" and consider the zero reference for evaluating spatial changes in equation (37) and we just analyze the variation of $V$ at a different moment of time.

Also, the voltage mentioned in equation (37) is actually the potential difference between the surface potential of the neuron membrane and the ground potential (taken as 0 V). Moreover, we are interested in the change in voltage and not the absolute value. So, we can use V in the equation without having to keep including the ground reference voltage.

## 5. Model validation

For an isolated neuron, we can get the model equation by taking p(x,t) = 0 in equation (37). This is because we are assuming that the neuron does not interact with other neurons i.e. it is isolated. So, we have:

$$V(x,t) = V_0 - \left(\frac{D}{\tau}\right)^{\frac{1}{2}} \int_0^t \frac{\partial V}{\partial x} \mathfrak{I}_0\left(\frac{t-t^*}{2\tau}\right) \exp\left(-\frac{t-t^*}{2\tau}\right) dt^* \quad (38)$$

The term $\frac{\partial V}{\partial x}$ on the right hand side of equation (38) is a direct result of the external influence i.e. stimulus. It is related to the flux as $\frac{\partial V}{\partial x} = -\frac{1}{g}\emptyset$ where g is the conductance and $\emptyset$ is the flux that can be defined by Gaussian distribution:

$$\emptyset = A\exp\left[-\frac{(t-t_{max})^2}{\delta^2}\right] \quad (39)$$

where $t_{max}$ denotes the moment of time, at which the Gaussian pulse (external stimulus) reaches its maximal value of $A$, whereas $\delta$ is the standard deviation.

This changes equation (38) to:

$$V(x,t) = V_0 + \frac{1}{g}\left(\frac{D}{\tau}\right)^{\frac{1}{2}} \int_0^t \emptyset \, \mathfrak{I}_0\left(\frac{t-t^*}{2\tau}\right) \exp\left(-\frac{t-t^*}{2\tau}\right) dt^* \quad (40)$$

Now, considering the effect of both sodium (Na) and potassium (K) as the driving ions for the action potential:

$$V(x,t) =$$
$$V_0 + \frac{1}{g_{Na}}\left(\frac{D_{Na}}{\tau_{Na}}\right)^{\frac{1}{2}} \int_0^t \emptyset_{Na} \, \mathfrak{I}_0\left(\frac{t_{Na}-t^*}{2\tau_{Na}}\right) \exp\left(-\frac{t_{Na}-t^*}{2\tau_{Na}}\right) dt^* + \frac{1}{g_K}\left(\frac{D_K}{\tau_K}\right)^{\frac{1}{2}} \int_0^t \emptyset_K \, \mathfrak{I}_0\left(\frac{t_K-t^*}{2\tau_K}\right) \exp\left(-\frac{t_K-t^*}{2\tau_K}\right) dt^* \quad (41)$$

The term $\frac{D}{\tau}$ is equal to $c^2$ where c is the velocity of propagation (21.2 $m/s$). The terms $\emptyset_{Na}$ and $\emptyset_K$, the flux of sodium and potassium respectively. So, the term $\emptyset_{Na}$ can be taken as:

$$\emptyset_{Na} = A_{Na} \exp\left[-\frac{(t-t_{max,Na})^2}{\delta_{Na}^2}\right] \quad (42)$$

and the term $\emptyset_K$ can be taken as:

$$\emptyset_K = A_K \exp\left[-\frac{(t-t_{max,K})^2}{\delta_K^2}\right] \quad (43)$$

where $A_{Na}$ is the maximum amplitude of the sodium flux, t is the total time, $A_K$ is the maximum amplitude of the potassium flux, t is the total time, $t_{max,Na}$ is the time of maximum flux for sodium, $t_{max,K}$ is the time of maximum flux for potassium, $\delta_{Na}$ is the standard deviation of Sodium flux and similarly $\delta_K$ for potassium flux.

The values to be substituted for the parameters involved are as given below in Table 1. It is noteworthy to mention that these values are governed from biological studies.

Table 1. Values of parameters to be substituted in equations (42), (43) and (44)

| Variable | Value | Units |
|---|---|---|
| $V_0$ | -0.07 | $V$ |
| c | 21.2 | $m/s$ |
| $g_{Na}$ | 1200 | $m^2/s$ |
| $g_K$ | 360 | $m^2/s$ |
| $A_{Na}$ | 5100 | $\frac{V-m}{s}$ |
| $A_K$ | 130 | $\frac{V-m}{s}$ |
| $t_{max,Na}$ | 0.0011 | s |
| $t_{max,K}$ | 0.002 | s |
| $\delta_{Na}$ | 0.0005 | s |
| $\delta_K$ | 0.001 | s |
| $\tau_{Na}$ | 0.0445 | s |
| $\tau_K$ | 0.002 | s |
| t | 0-0.005 | s |

In fact, these parameters represent the neuron firing characteristics as the result of receiving external stimuli which is equated with sodium and potassium flux using $\frac{\partial V}{\partial x} = -\frac{1}{g}\emptyset$.

The sodium flux can be plotted from equation (42) as given in Figure. 3 below:

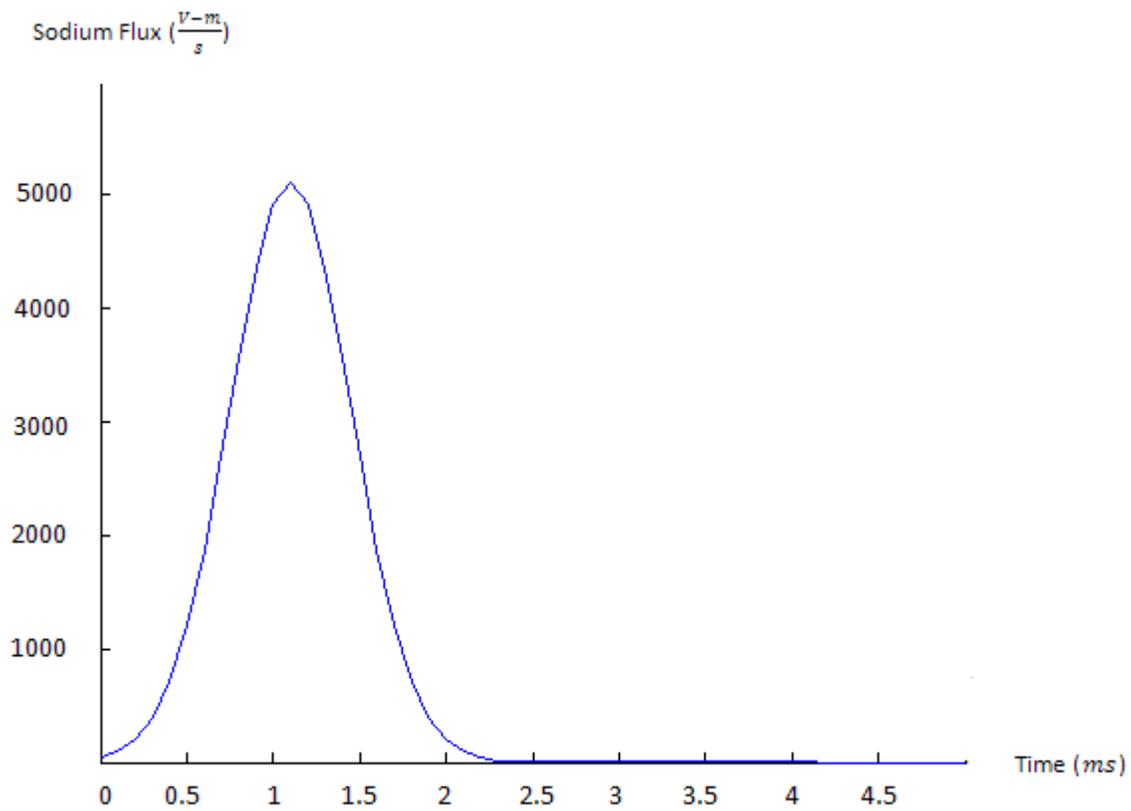

Figure. 3. Sodium flux as plotted from equation (42). The flux rises to a maximum value around 1 ms as expected and then goes down to $0 \frac{V-m}{s}$.

The potassium flux can be plotted from equation (43) as given in Figure. 4 below:

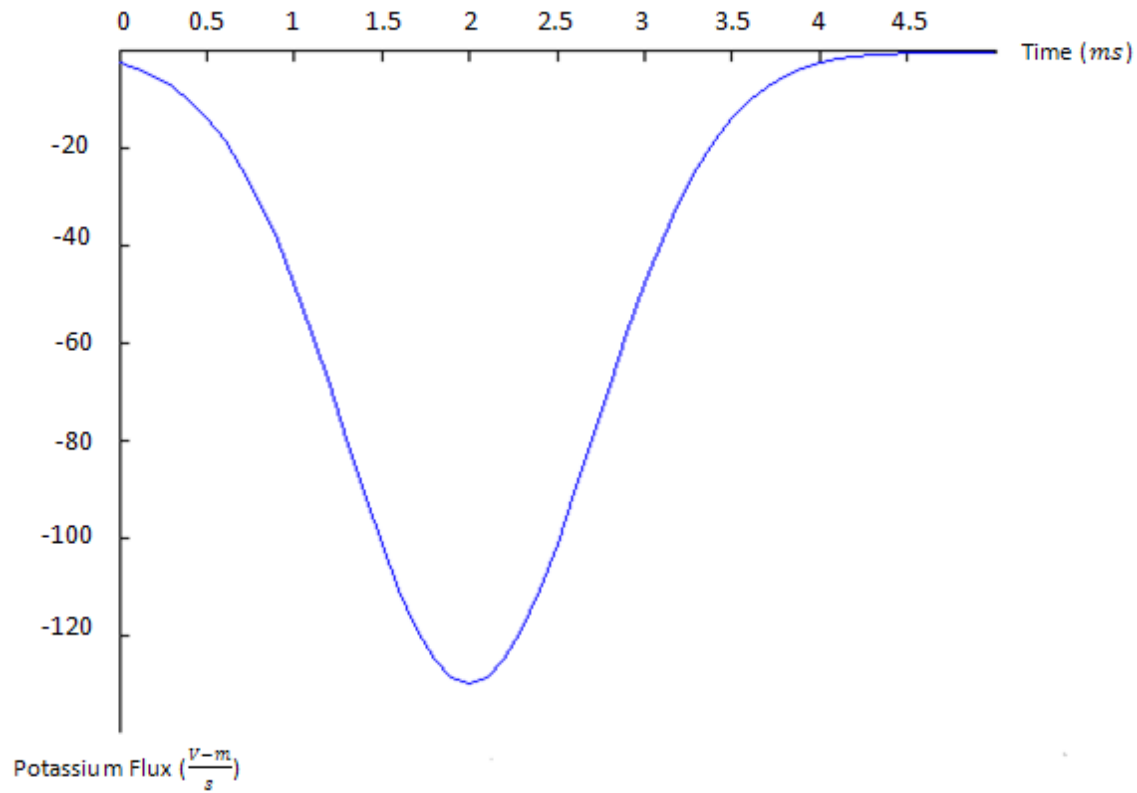

Figure. 4. Potassium flux as plotted from equation (43). The flux slowly rises to a maximum value around 2 ms and then gradually goes down to $0 \frac{V-m}{s}$.

It can be seen in Figure.3 that the sodium flux rises to its maximum value near t = 0.0011 sec after which it reduces to $0 \frac{V-m}{s}$. On the other hand, the potassium flux slowly rises and reaches its maximum value near t = 0.002 sec after which it reduces to $0 \frac{V-m}{s}$. This is the behavior that had been described by various researchers when they performed experiments on the giant axon of the squid [23-26]. The opening and closing of sodium and potassium channels lead to this variation in flux.

Substituting the equation (42), and (43) in equation (41):

$$V(x,t) = V_0 + \frac{c}{g_{Na}} \int_0^t A_{Na} \exp\left[-\frac{(t-t_{max,Na})^2}{\delta_{Na}^2}\right] \mathcal{I}_0\left(\frac{t_{Na}-t^*}{2\tau_{Na}}\right) \exp\left(-\frac{t_{Na}-t^*}{2\tau_{Na}}\right) dt^* +$$
$$\frac{c}{g_K} \int_0^t A_K \exp\left[-\frac{(t-t_{max,K})^2}{\delta_K^2}\right] \mathcal{I}_0\left(\frac{t_K-t^*}{2\tau_K}\right) \exp\left(-\frac{t_K-t^*}{2\tau_K}\right) dt^* \quad (44)$$

The equation (44) consists of the voltages for both sodium and potassium fluxes. When the equation (44) is solved and plotted, we get the plot as shown in Figure. 5.

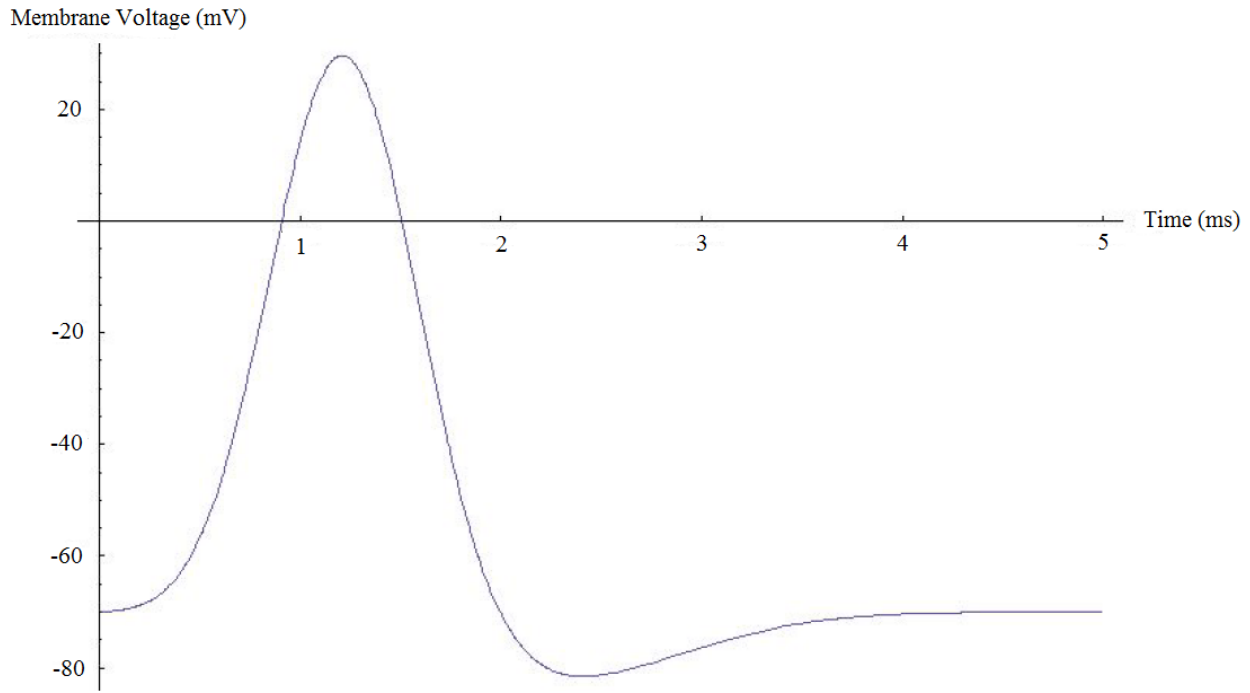

Figure. 5. Solution of equation (44) - A plot similar to action potential is generated.

Figure. 5 shows the characteristic shape and values of the action potential. The resting potential has been taken as -70 mV. The potential rises to a maximum of + 30 mV before it falls down to around -80 mV and subsequently rising again to reach the resting potential value of -70 mV. This is due to the fact that sodium channels rapidly open up when the action potential is initiated leading to a sodium ion influx till the transmembrane voltage approaches a value closer to the sodium Nernst potential. The sodium channels then close and the potassium channels open to cause an efflux of potassium ions, which continues till the potential approaches the potassium Nernst potential. These causes a slight undershoot which is also called hyperpolarization. Later the ionic balance is restored which leads to the transmembrane potential become closer to the resting potential value.

The generated action potential plot is similar to the real action potential plot which is shown in Figure 2. Thus, we can say that using the proposed mathematical equations we modeled the generation of single action potential in a node of Ranvier. It is noteworthy to mention that this model can be developed in case of neuronal interactions. If so, we consider the source function in equation (28) which bring us a general model of neural activity in mesoscopic and further in macroscopic level of brain organization.

## 6. Conclusion

The proposed model incorporates the concept of finite propagation speed of the action potential in the neuron by introducing a finite time lag that eliminates the paradox of instantaneous propagation which is present in the diffusion model. Solving this model gives a relation between the voltage in a single neuron to its spatial derivative at any given moment of time and location. On the application of the required values of the parameters involved, a plot of a single action potential is generated by the model which is similar to the real action potential plot generated by biologist. The plots of sodium and potassium fluxes as provided by the model are consistent with the general behavior of the ions inside the neuron. This model shall be further applied in the case of neuronal interactions and the results verified against currently available EEG records.